\documentclass[aps,reprint,amsmath, amssymb,groupedaddress,floatfix]{revtex4-1}
\usepackage{natbib}
\usepackage{graphicx}
\usepackage{bm}
\usepackage{color,soul}
\usepackage{hyperref}% add hypertext capabilities
\usepackage{float}
\usepackage{csquotes}
\usepackage{color,soul}
\usepackage{hyperref}% add hypertext capabilities
\hypersetup{colorlinks=true, urlcolor=blue, citecolor=blue}
\usepackage{longtable}
\usepackage{tabularx}

\begin{document}

\title{Large electrically and chemically tunable Rashba-Dresselhaus effects in Ferroelectric CsGeX$_3$ (X=Cl, Br, I) perovskites}
\author{Abduljelili Popoola}
\email{Popoola@usf.edu}
\author{Nikhilesh Maity}
\author{Ravi Kashikar}
\author{S. Lisenkov}
\author{I. Ponomareva}
\email{iponomar@usf.edu}
\affiliation{Department of Physics, University of South Florida, Tampa, Florida 33620, USA}

\date{\today}

\begin{abstract}

Rashba-Dresselhaus effects, which originate from spin-orbit coupling and allow for spin manipulations, are actively explored in materials following the pursuit of spintronics and quantum computing. However, materials that possess practically significant Rashba-Dresselhaus effects often contain toxic elements and offer little opportunity for tunability of the effects. We used first-principles simulations to reveal that the recently discovered halide ferroelectrics in the CsGeX$_3$ (X=Cl, Br, I) family possess large and tunable Rashba-Dresselhaus effects. In particular, they give origin to the spin splitting of up to 171~meV in valence band of CsGeI$_3$. The value is chemically tunable and can decrease by 25\% and 70\% for CsGeBr$_3$ and CsGeCl$_3$, respectively. Such chemical tunability could result in engineering of desired values through solid solution technique. Application of electric field was found to result in structural changes that could both decrease and increase spin splitting leading to electrical tunability of the effect. In the vicinity of conduction and valence band extrema, the spin textures are mostly of Rashba type which is promising for spin-to-charge conversion applications. The spin directions are coupled with the polarization direction leading to Rashba-ferroelectricity co-functionality. Our work identifies lead-free perovskite halides as excellent candidates for spin-based applications and is likely to stimulate further research in this direction.

\end{abstract}
\maketitle

Germanium based halide perovskites CsGeX$_3$ (X=Cl, Br, I) stand out among other inorganic halide perovskites, thanks to their unique features. Firstly, these materials crystallize in polar phases at room temperature and exhibit single phase transition from cubic to polar rhombohedral phase~\cite{CGX_exp1}. Secondly, their spontaneous polarization is comparable to that found in canonical ferroelectric oxides~\cite{CGX_exp2, CGX_exp3, Database}. Additionally, atomistic insight from the effective Hamiltonian simulations  predicted that, these materials possess competing polar and antipolar phases which result in polarization fields and domain configurations which are energetically prohibited in canonical ferroelectric oxides~\cite{CNG6}.

One of the consequences of broken inversion symmetry in these materials is the appearance of a local electric field, which couples with the electron spin and momentum to give origin to the spin-orbit-coupling (SOC) which selectively lifts spin degeneracy in momentum space~\cite{Kramer}. This is known as Rashba~\cite{Rashba} or Dresselhaus~\cite{Dresselhaus} effects. Thus, these materials could have potential for applications in spintronics: the field that exploits an electron's spin, rather than its charge~\cite{spintronics}. In addition to being ferroelectric, these materials also exhibit dispersive bands with electronic bandgap ranging from 1.6 to 3.4~eV~\cite{PCE_CGX2, Eb5}, revealing their potential for multifunctional applications. However, currently, there exists a gap in our understanding of the Rashba-Dresselhaus effects in these materials and their coupling with ferroelectricity. Many questions remain unanswered. For example, what is the magnitude of spin-splitting in these materials and how does it change across different members of the family? Is this variation strong enough to allow for tunability of the spin splitting through chemical substitution? What type of spin textures does the family posses in the vicinity of the conduction band minimum (CBM) and valence band maximum (VBM) and what potential applications could it allow? How tunable are these electronic structure effects by structural modifications caused by an electric field? Finally, can spin textures be manipulated or tuned by both chemical substitution on the halogen site and electric field? 

The electronic structure of CsGeX$_3$ has been computationally studied in Refs.~\cite{Eg_CGX1, Eg_CGX2, Eg_CGX3, Eg_CGX4} where it was found that orbital character of VBM are dominated by X-5p and Ge-4s orbitals whereas Ge-4p orbitals show dominance at the CBM. Within density functional theory (DFT) calculations, electronic bandgap of CsGeX$_3$ were predicted to be in range of 0.6-1.4~eV \cite{Eb5, Eg_CGX4, Eg_CGX5} which are underestimated with respect to experimental values. Later on, the quasiparticle-based GW formalism was used to improve the bandgap values~\cite{Eb2, Eb3, Eb5}. However, the effect of spin orbit coupling (SOC) was not considered in these calculations. In Refs.~\cite{WhenSOChelps, PhysRevB.93.195211}, it was shown that incorporating SOC slightly lowers the electronic bandgap of CsGeX$_3$; demonstrating a potential for spin splitting. Within high-throughput computational screening of materials, it was shown that CsGeI$_3$ exhibit Rashba spin splitting of 26~meV~\cite{Zunger4}. However, a comprehensive analysis of Rashba-Dresselhaus effects in CsGeX$_3$ is presently missing. In this work, we aim to fill up the gap in the knowledge of Rashba-Dresselhaus effects and their coupling to ferroelectricity in CsGeX$_3$ family using density functional theory (DFT) calculations. In particular, our goals are to: (i) predict the existence of large spin splitting, spin textures and Rashba-ferroelectricity co-functionality in these materials; (ii)  report chemical tunbaility of these properties as X-site anion changes from Cl$^{-}$ to Br$^{-}$ and then to I$^{-}$; (iii) provide an insight into tunability of these properties by electric field.

To achieve our goals, we perform DFT simulations using Vienna ab-initio simulation package (VASP)~\cite{VASP1, VASP2, VASP3, VASP4}. We utilized Perdew-Burke-Ernzerhof's (PBE)~\cite{PBE} version of generalized gradient approximation within the projector augmented wave (PAW) pseudopotential~\cite{blochl94} for our calculations. We initiated our calculations with experimental structures as reported in Refs~\cite{CGX_exp1,CGX_exp2,CGX_exp3} and subjected the structures to relaxations using conjugate gradient algorithm until the ionic forces was below 1~meV/\AA. We set energy cutoff for plane wave basis at 600~eV and used a $\Gamma$-centered Monkhorst-Pack k-point mesh of 10 × 10 × 8 to perform integration within the Brillouin zone~\cite{VASPKIT, KPOINTS}. We performed electronic structure calculations with and without SOC using the aforementioned parameters. For spin texture calculation, we used a 25 × 25 k-point mesh for a specified plane in the reciprocal space as generated by PYPROCAR~\cite{pyprocar}. For polarization calculation, we used the modern theory of polarization developed by King-Smith and Vanderbilt~\cite{Berry_phase}.

\begin{figure*}[h]
\centering
\includegraphics[width=\textwidth]{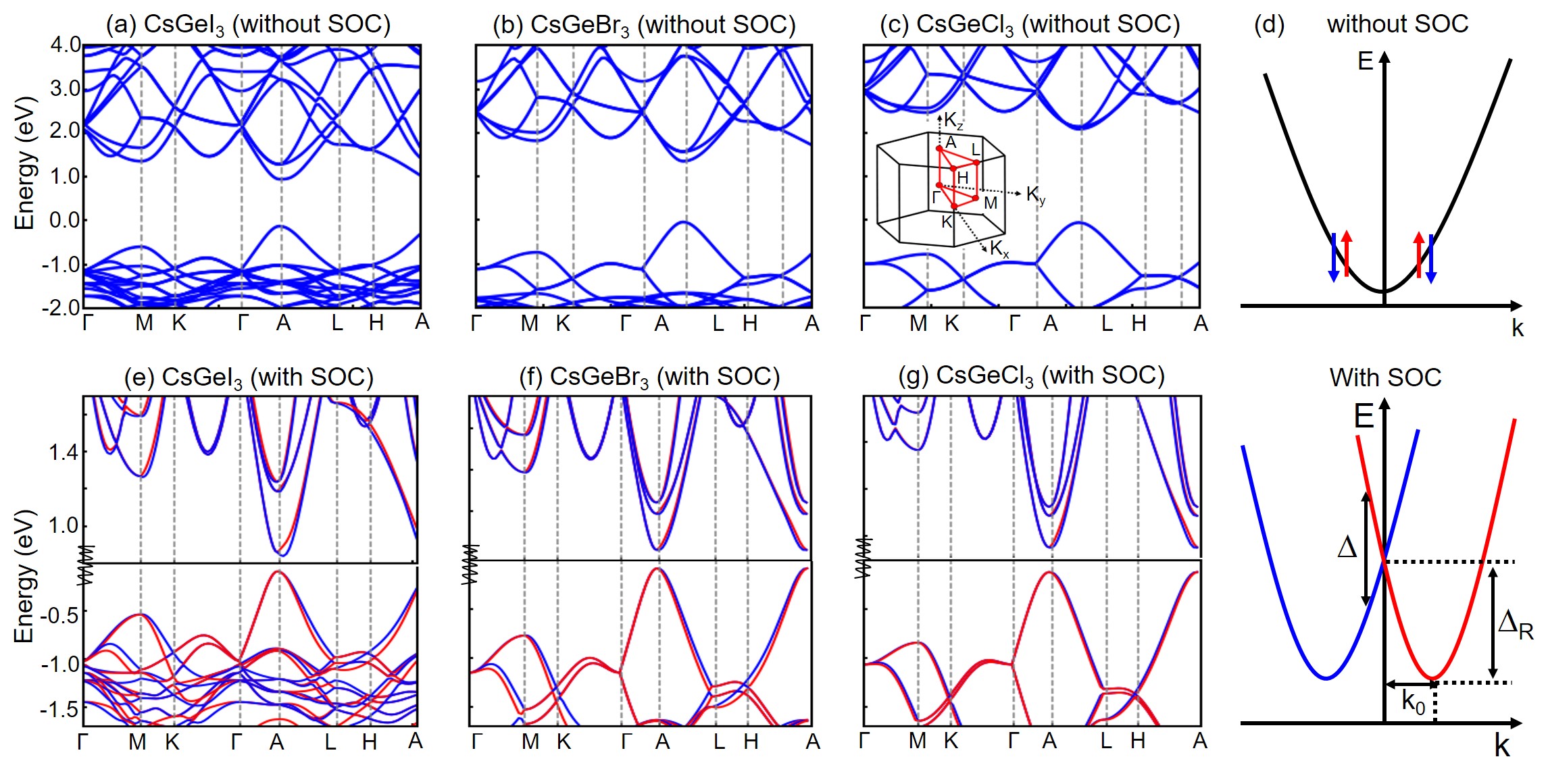}
\caption{Electronic band structures of CsGeX$_3$ (a)-(c) without and (e)-(g) with SOC (d) schematic diagram indicating $\Delta$, $\Delta_{R}$ and $k_0$}
\label{Fig2}
\end{figure*}

We begin with the investigation of electronic structure of CsGeX$_3$. Figures~\ref{Fig2}(a)-(c) show the band dispersion computed without SOC. The data indicate that all materials have CBM and VBM at point A (0.0, 0.0, 0.5) in the Brillouin zone which is schematized in the inset of Figure~\ref{Fig2}(c). The electronic bandgaps are 1.07~eV, 1.39~eV and 2.14~eV for CsGeI$_3$, CsGeBr$_3$ and CsGeCl$_3$, respectively. These values are underestimated relative to experimental values, as expected for the choice of exchange correlation functional~\cite{Eg_gap}, but are comparable with previously reported values~\cite{WhenSOChelps, PCE_CGX2, Eb5}. The analysis of partial density of states shows that the bands around the Fermi level are dominated by Ge-\{s, p\} and (Cl, Br, I)-p orbitals. Specifically, the contribution near conduction band minimum (CBM) is dominated by Ge-p orbitals with values in range 72-77\%. There are also small contributions from Ge-s and (Cl, Br, I)-\{s, p\} with values in range 0-14\% and 13-21\% respectively. Near the valence band maximum (VBM), Ge-s and (Cl, Br, I)-p orbitals show dominance with contributions in range 31-36\% and 54-61\% respectively. Ge-p orbitals also show small contributions in range 1-13\%. Figures~\ref{Fig2}(e)-(g) show the band structure computed in the presence of SOC. Now the bands are splitted, including in the vicinity of CBM and VBM, signifying presence of Rashba and/or Dressalhaus effects. 
To quantify the effects, we will use the following metrics: (i) momentum-dependent spin splitting, $\Delta$, defined as the difference in energy eigenvalues of spin-splitted bands at the given point in the momentum space; (ii) Rashba spin splitting, $\Delta_{R}$, defined as the difference in energy eigenvalues of degenerate and non-degenerate extrema, typically around the CBM and VBM (see Figure~\ref{Fig2}(d)); (iii) Rashba coefficient, defined as  $\alpha_{R} = \frac{2\Delta_{R}}{k_0}$; where $k_0$ is  offset away from the degenerate extrema (see Figure~\ref{Fig2}(d)). It should be noted that the symmetry of the R3m phase (C$_{3v}$ point group) allows for Rashba and Dresselhaus effects~\cite{Zunger1}. The A point in the Brillouin zone, which hosts CBM and VBM, has C$_{3v}$ wave vector point group, thus allowing for Rashba effects only~\cite{Zunger1}. Therefore, computing Rashba figures of merit are appropriate in this case.

\begin{figure*}[h]
\centering
\includegraphics[width=0.75\textwidth]{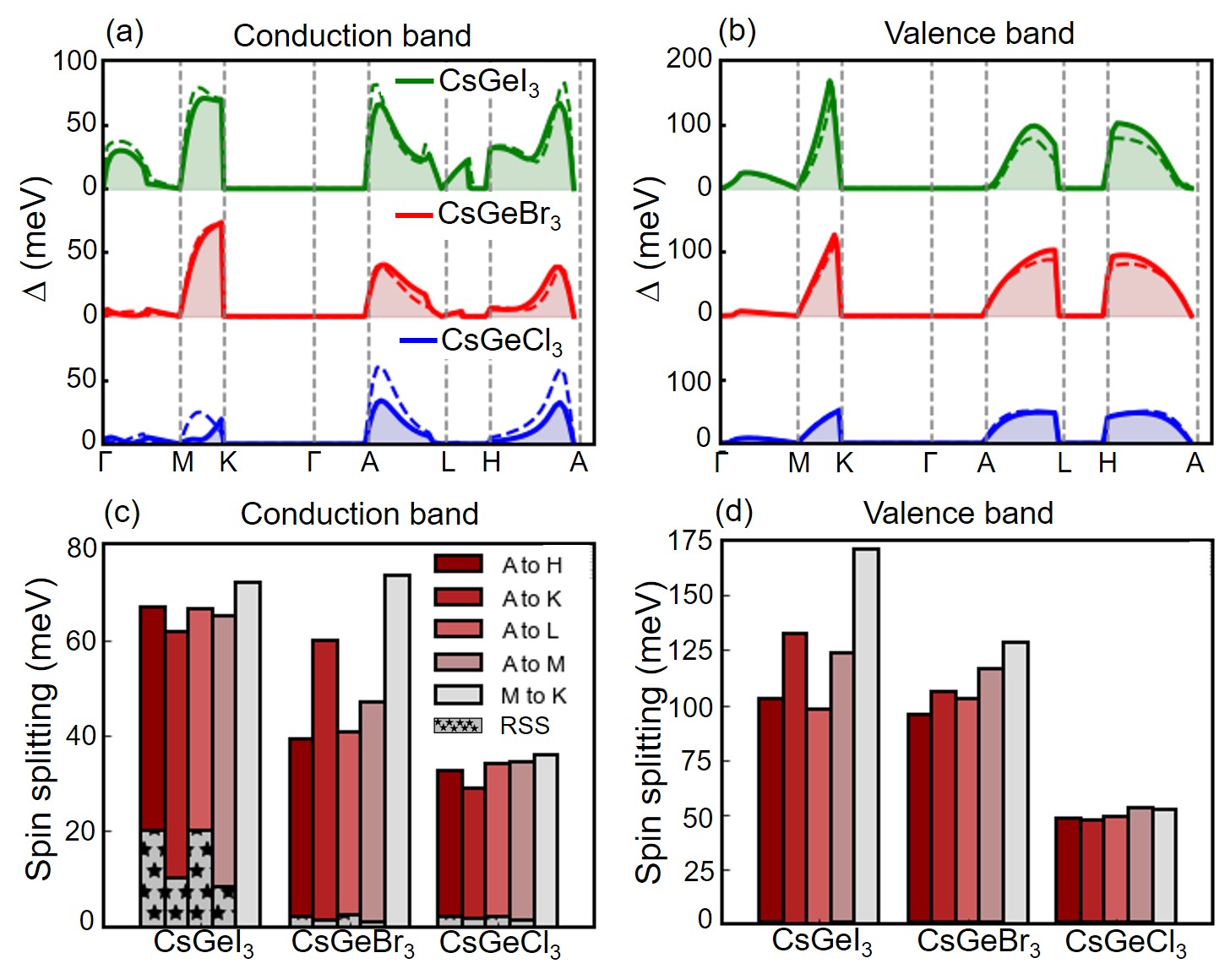}
\caption{(a)-(b) The conduction and valence band $\Delta$ along the high symmetry directions in the Brillouin zone. Dashed lines correspond to the spin splitting at finite temperature of 300~K, (c)-(d) The maximum $\Delta$ and $\Delta_R$ near band edge along different symmetry directions in the Brillouin zone.}
\label{Fig3}
\end{figure*}

\begin{figure*}[h]
\centering
\includegraphics[width=0.4\textwidth]{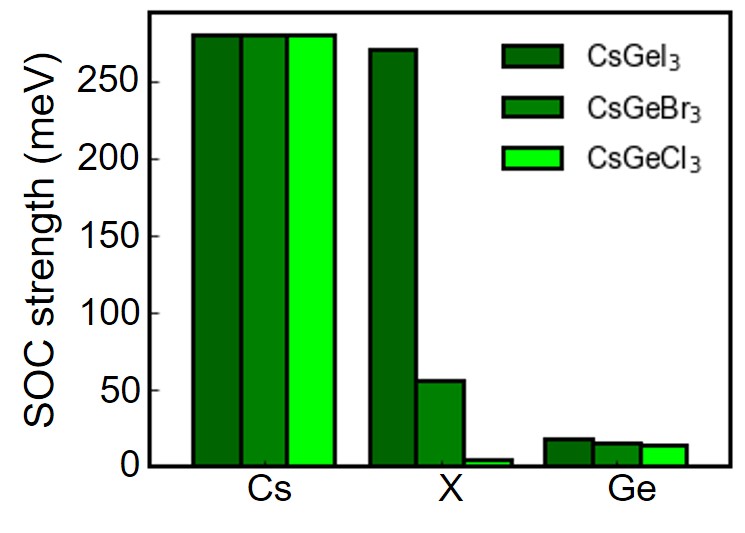}
\caption{The effective SOC energies for Cs, Ge, and X (= I, Br, Cl) atoms}
\label{Fig3a}
\end{figure*}

Figures~\ref{Fig3}(a)-(b) show $\Delta$ along some select paths in the Brilloun zone. We find that all three materials exhibit similar type of dependencies. Nevertheless, there exist significant quantitative differences, which can be better assessed from Figure~\ref{Fig3}(c)-(d), which reports the maximum values along a few chosen directions in the Brillouin zone. The data reveal that, in the conduction band, $\Delta$ values for CsGeI$_3$ can exceed those for CsGeBr$_3$ and CsGeCl$_3$ by 70\% and 25\% respectively. In the valence band, CsGeI$_3$ and CsGeBr$_3$ demonstrate comparable values, which are twice the ones found for CsGeCl$_3$. These findings of difference in spin splitting in CsGeI$_3$, CsGeBr$_3$ and CsGeCl$_3$ signify large chemical tunability of Rashba and/or Dressalhaus effects in these materials. In the valence band, $\Delta$ can reach a very large value of 171~meV, which is well above thermal energy at room temperature, and therefore, is of practical importance. In order to understand these difference, we have analyzed the SOC strength of each element and report it in Figure~\ref{Fig3a}. The contribution from Cs is constant for all the three materials. Ge SOC is also comparable between the different materials. The large difference is present for the halide element, which explains why CsGeI$_3$ exhibits largest spin-splitting among different materials. 

\begin{table*}[!h]
    \centering
    \footnotesize
    \caption{Rashba parameters for CsGeX$_3$: $\Delta_{R}$, $k_0$ and $\alpha_{R}$ calculated near CBM and VBM along $A \rightarrow L$ in the Brillouin zone. VBM values are in the parenthesis}
    \begin{ruledtabular}
    \begin{tabular}{cccc}
         material & $\Delta_{R}$ (meV) & $k_0$ (\AA$^{-1}$) & $\alpha_{R}$ (meV\AA)\\
         \hline
         CsGeI$_3$ & 20.10 (0.00) & 0.043 (--)     & 934.42 (0.0)\\
         CsGeBr$_3$ & 2.40 (0.80) & 0.018 (0.0091) & 262.84 (175.22) \\
         CsGeCl$_3$ & 2.20 (0.50) & 0.019 (0.0094) & 233.05 (105.93)\\
    \end{tabular}
    \end{ruledtabular}
    \label{Table1}
\end{table*}

Rashba spin-splitting in the vicinity of the A point, along $A \rightarrow L$, is also reported in Figure~\ref{Fig3}(c). For all the materials, $\Delta_{R}$ near the VBM is small. However, near the CBM, $\Delta_{R}$ is significant and largest for CsGeI$_3$ as presented in Table~\ref{Table1}. We find $\Delta_{R}$ for CsGeI$_3$ to be comparable with previous high-throughput study~\cite{Zunger4}. The calculated values of $\Delta_{R}$ and other Rashba figures of merits ($k_0$ and $\alpha_{R}$) are presented in Table~\ref{Table1}. CsGeI$_3$ values are comparable with conduction band Rashba parameters of other bulk crystals including AuCN, LiSbZn, IrSnS among others~\cite{Zunger4, Picozzi}. In particular, $\alpha_{R}$ narrowly classifies CsGeI$_3$ as strong Rashba compound. CsGeCl$_3$ and CsGeBr$_3$ classify as weak Rashba compounds~\cite{Zunger4}. To gain some insights into the role of temperature, we carried out simulations on the structure in which lattice constants are kept the same as in experimental structure collected at 298~K~\cite{CGX_exp1}. The experimental values of the hexagonal lattice parameters are: a = b = 8.36~\AA, c = 10.61~\AA, for CsGeI$_3$; a = b = 7.88~\AA, c = 9.97~\AA, for CsGeBr$_3$; a = b = 7.67~\AA, c = 9.46~\AA, for CsGeCl$_3$. The spin splitting, given by dashed lines shown in Figures~\ref{Fig3}(a)-(b), and Rashba figures of merit indicate no significance difference. This suggests that our values provide good estimate for the room temperature ones. 
 
\begin{figure*}[h]
\centering
\includegraphics[width=\textwidth]{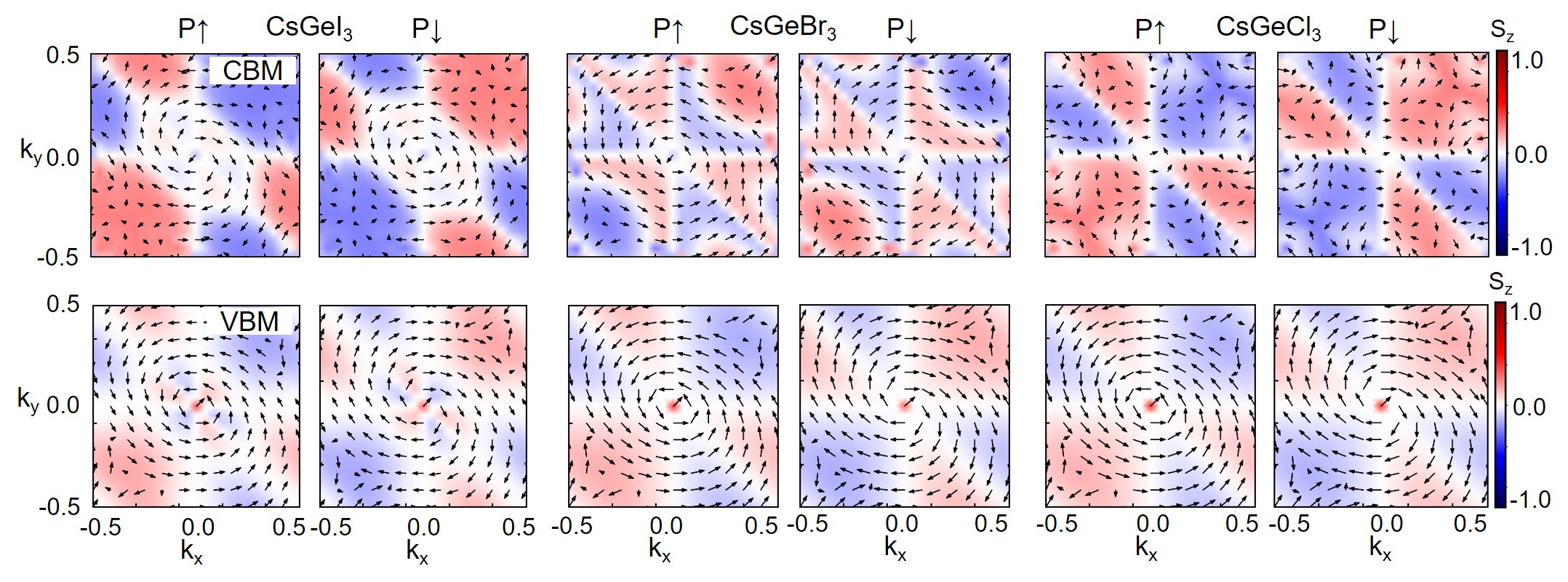}
\caption{Near band edge spin textures of CsGeX$_3$ for different directions of polarization as given in the titles}
\label{Fig4}
\end{figure*}

The spin textures in the reciprocal space, which are the expectation values of the spin operator, are presented for the (k$_x$, k$_y$, 0.5) plane in Figure~\ref{Fig4}. This is the plane perpendicular to the polarization. The spin textures are reported for the vicinity of VBM and CBM. As already mentioned, the spin textures in the vicinity of A point are expected to be of Rashba type which is consistent with symmetry considerations~\cite{Zunger1}. Indeed, this is the case in our calculations. Rashba type of spin textures are useful for applications in spin-to-charge conversion~\cite{Jagoda1, REE2}. As we move away from the A point, the type of spin textures changes and becomes a mixture of Rashba and Dresselhaus. The reason is that the point group of the wave vector changes from C$_{3v}$ to C$_s$, thus allowing to co-host these two types of spin textures. As shown in Figure~\ref{Fig4}, reversal of polarization direction results in reversal of the direction for the spins expectation values, demonstrating the concept of ferroelectric-Rashba co-functionality~\cite{CNG3,CNG1}.

\begin{figure*}[h]
\centering
\includegraphics[width=\textwidth]{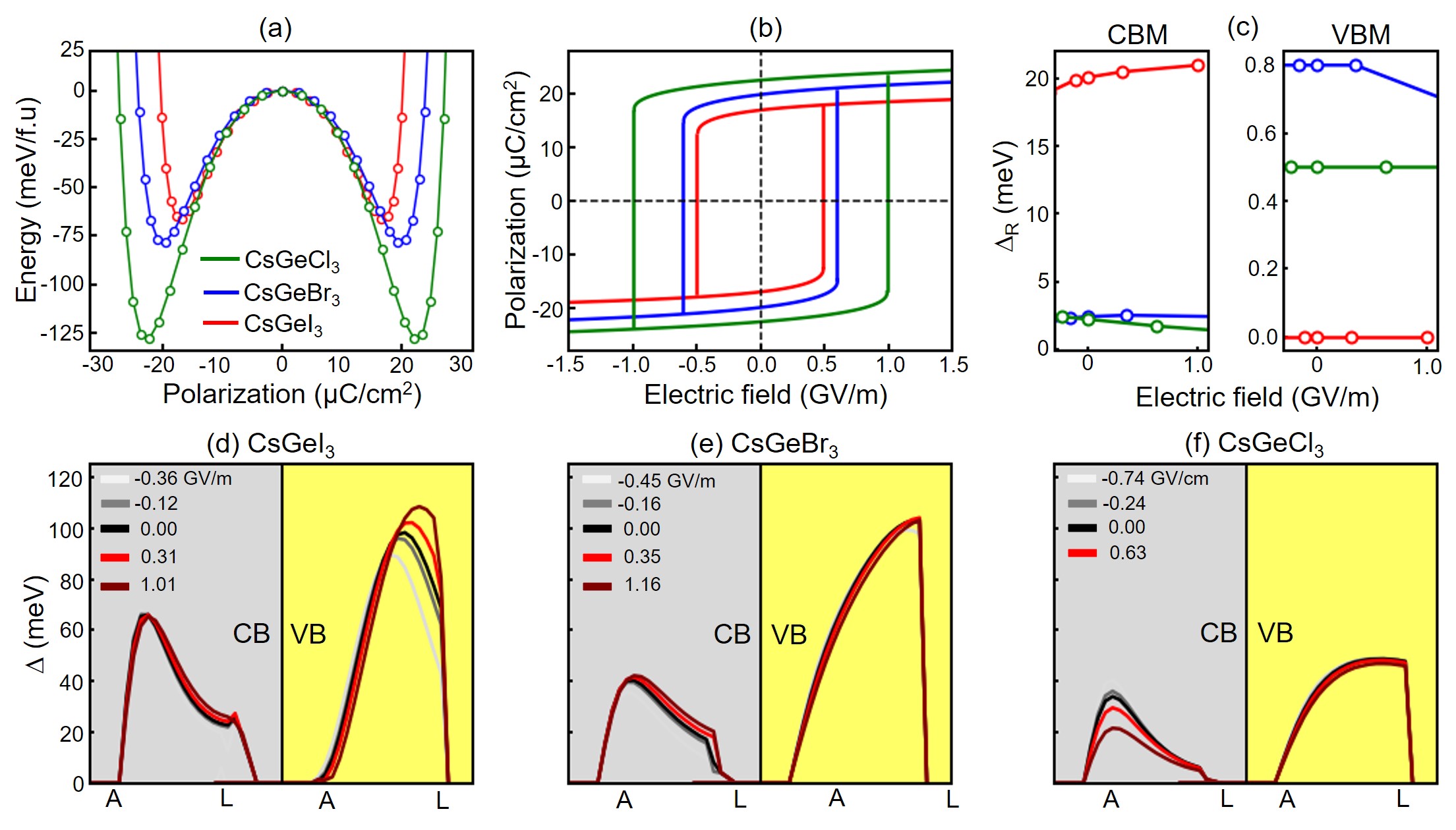}
\caption{(a) Supercell energy as a function of polarization (b) P-E loop (c) variation of $\Delta_R$ with electric field for CsGeX$_3$. (d)-(f) Spin splitting along A-L for CsGeX$_3$ for different values of electric field, as indicated in the titles. The grey and yellow regions represent the conduction and valence bands respectively}
\label{Fig5}
\end{figure*}

Next we aim to investigate tunability of the spin splitting and spin textures by the electric field. For that, we first compute energy along the polarization reversal path (see Figure~\ref{Fig5}(a)). The energy profile is then fitted with the equation $F(P) = aP^2 + bP^4 + cP^6 - EPV$, where $a$, $b$, $c$ are the coefficients in the free energy expansion and V is the volume of supercell. This equation is then used to compute the equation of state from $\frac{\partial F}{\partial P}=0$. This yields P-E loops shown in  Figure~\ref{Fig5}(b) which reveal chemical tunability of both spontaneous polarization and coercive field.  
Next, $\Delta$ is computed along $A \rightarrow L$ for a few points on the upper branch of the hysteresis loop, that is in the interval [-0.4:1.2]~GV/m. Figures~\ref{Fig5}(d)-(f) show that $\Delta$ increases with electric field in the conduction and valence bands of CsGeI$_3$ and CsGeBr$_3$. In particular, maximum $\Delta$ along $A \rightarrow L$ increases from 98~meV to 109~meV at the valence band of CsGeI$_3$. For CsGeCl$_3$, $\Delta$ shows very weak tunability. Figure~\ref{Fig5}(c) shows an evolution of $\Delta_R$ around A point with electric field for both conduction band and valence band. For conduction band, $\Delta_R$ increases very slightly with electric field, by 1.1~meV under 1~GV/m of electric field. Whereas, for valence band, $\Delta_R$ remains constant in this electric field regime. For CsGeBr$_3$ and CsGeCl$_3$, $\Delta_R$ show similar trend with electric field.

\begin{figure*}[h]
\centering
\includegraphics[width=\textwidth]{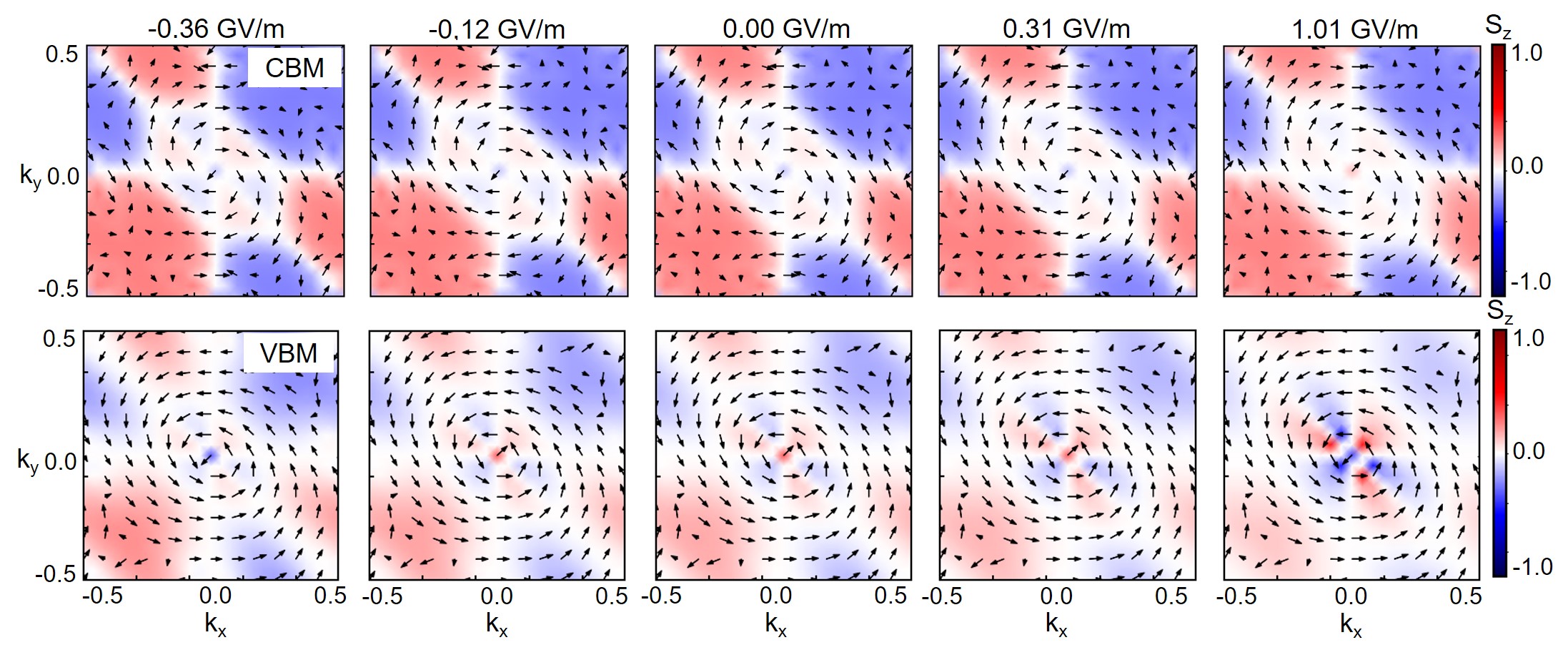}
\caption{Variation of near band edge spin texture of CsGeI$_3$ with electric field.}
\label{Fig6}
\end{figure*}

Figure~\ref{Fig6} shows evolution of spin textures around CBM and VBM with an electric field for CsGeI$_3$. In the conduction band, we notice that the area accommodating Rashba type spin textures increases with the electric field. The same effect is a bit less pronounced in the VB as large portion of the Brilloin zone is already occupied with Rashba type spin textures. Similar effects were observed for CsGeBr$_3$ and CsGeCl$_3$. 

In summary, we have used DFT simulations to predict Rashba-Dresselhaus effects in the family of ferroelectric halide perovskites CsGeX$_3$ (X= I, Br, Cl) and their coupling with ferroelectricity. We found that SOC results in spin-splitting in both conduction and valence band of these materials. Among the three materials in the family, CsGeI$_3$ exhibits the largest spin splitting in both valence and conduction band. The largest value can reach 171~meV opening opportunity for potential applications in spintronics. We found no spin splitting along the polar direction. The three materials exhibit rather similar trends for the spin-splitting evolution in the Brillouin zone, however, there are quantitative differences which were found to originate from atomic SOC of the halides. Owing to the wave vector point group of the CBM and VBM all the materials exhibit Rashba spin textures near the band edges. The spin direction is coupled to the direction of electric polarization and can be reversed, which is known as Rashba ferroelectricity co-functionality. Comparison between the computational data obtained for the ground state and room temperature for some instances reveal little differences, suggesting that our results can be extrapolated to room temperatures. Application of electric field induces structural changes, which were computed from the equation of state. Such changes, in turn, result in the changes in spin-splitting and spin textures. For example, along A to L, spin splitting can increase from 96~meV to 109~meV in CsGeI$_3$ through the application of electric field of 1.01~GV/m. Application of electric field in the direction opposite to polarization can decrease this value down to 90~meV through the application of electric field of $-$0.36~GV/m. Furthermore, application of electric field along the direction of polarization converts spin textures into Rashba type so that a larger portion of the Brillouin zone becomes occupied with such type of pattern. We note that pure Rashba types of spin textures are valuable for spin-charge conversion. We believe that our work provides an important insight into multifucntional properties of CsGeX$_3$ material and will promote both further investigation and technological applications. 

\subsection{Acknowledgment}

This work was supported by the U.S.
Department of Energy, Office of Basic Energy Sciences, Division of Materials Sciences and Engineering under Grant No. DE-SC0005245.  Computational support was provided by the National Energy Research Scientific Computing Center (NERSC), a U.S. Department of Energy, Office of Science User Facility located at Lawrence Berkeley National Laboratory, operated under Contract No. DE-AC02-05CH11231 using NERSC award BES-ERCAP-0025236.


\begin{thebibliography}{10}

\bibitem{CGX_exp1}
G.~Thiele, H.~W. Rotter, and K.~D. Schmidt, ``Kristallstrukturen und
  phasentransformationen von caesiumtrihalogenogermanaten(ii) csgex3 (x = cl,
  br, i),'' {\em Zeitschrift für anorganische und allgemeine Chemie},
  vol.~545, no.~2, pp.~148--156, 1987.

\bibitem{CGX_exp2}
R.~Chen, C.~Liu, Y.~Chen, C.~Ye, S.~Chen, J.~Cheng, S.~Cao, S.~Wang, A.~Cui,
  Z.~Hu, H.~Lin, J.~Wu, X.~Y. Kong, and W.~Ren, ``Ferroelectric csgei3 single
  crystals with a perovskite structure grown from aqueous solution,'' {\em The
  Journal of Physical Chemistry C}, vol.~127, no.~1, pp.~635--641, 2023.

\bibitem{CGX_exp3}
Y.~Zhang, E.~Parsonnet, A.~Fernandez, S.~M. Griffin, H.~Huyan, C.-K. Lin,
  T.~Lei, J.~Jin, E.~S. Barnard, A.~Raja, P.~Behera, X.~Pan, R.~Ramesh, and
  P.~Yang, ``Ferroelectricity in a semiconducting all-inorganic halide
  perovskite,'' {\em Science Advances}, vol.~8, no.~6, p.~eabj5881, 2022.

\bibitem{Database}
T.~E. Smidt, S.~A. Mack, S.~E. Reyes-Lillo, A.~Jain, and J.~B. Neaton, ``An
  automatically curated first-principles database of ferroelectrics,'' {\em
  Scientific Data}, vol.~7, no.~1, pp.~2052--4463, 2020.

\bibitem{CNG6}
R.~Kashikar, S.~Lisenkov, and I.~Ponomareva, ``Coexistence of polar and
  antipolar phases in ferroelectric halide perovskite
  ${\mathrm{csgebr}}_{3}$,'' {\em Phys. Rev. B}, vol.~109, p.~L020101, Jan
  2024.

\bibitem{Kramer}
M.~J. Klein, ``{On a Degeneracy Theorem of Kramers},'' {\em American Journal of
  Physics}, vol.~20, pp.~65--71, 02 1952.

\bibitem{Rashba}
E.~RASHBA, ``Properties of semiconductors with an extremum loop. i. cyclotron
  and combinational resonance in a magnetic field perpendicular to the plane of
  the loop,'' {\em Sov. Phys.-Solid State}, vol.~2, p.~1109, 1960.

\bibitem{Dresselhaus}
G.~Dresselhaus, A.~F. Kip, and C.~Kittel, ``Spin-orbit interaction and the
  effective masses of holes in germanium,'' {\em Phys. Rev.}, vol.~95,
  pp.~568--569, Jul 1954.

\bibitem{spintronics}
A.~Hirohata, K.~Yamada, Y.~Nakatani, I.-L. Prejbeanu, B.~Diény, P.~Pirro, and
  B.~Hillebrands, ``Review on spintronics: Principles and device
  applications,'' {\em Journal of Magnetism and Magnetic Materials}, vol.~509,
  p.~166711, 2020.

\bibitem{PCE_CGX2}
L.-J. Chen, ``Synthesis and optical properties of lead-free cesium germanium
  halide perovskite quantum rods,'' {\em RSC Adv.}, vol.~8, pp.~18396--18399,
  2018.

\bibitem{Eb5}
N.~Thi~Han, V.~Khuong~Dien, and M.-F. Lin, ``Electronic and optical properties
  of csgex3 (x= cl, br, and i) compounds,'' {\em ACS Omega}, vol.~7, no.~29,
  pp.~25210--25218, 2022.

\bibitem{Eg_CGX1}
S.~Bouhmaidi, A.~Marjaoui, A.~Talbi, M.~Zanouni, K.~Nouneh, and L.~Setti, ``A
  dft study of electronic, optical and thermoelectric properties of ge-halide
  perovskites csgex3 (x=f, cl and br),'' {\em Computational Condensed Matter},
  vol.~31, p.~e00663, 2022.

\bibitem{Eg_CGX2}
S.~S.~I. Almishal and O.~Rashwan, ``A comparative study of the structural and
  electronic properties of orthorhombic and cubic cspbi3 and trigonal csgei3
  using first-principles calculations,'' in {\em 2021 IEEE 48th Photovoltaic
  Specialists Conference (PVSC)}, pp.~1837--1841, 2021.

\bibitem{Eg_CGX3}
D.~Ray, C.~Clark, H.~Q. Pham, J.~Borycz, R.~J. Holmes, E.~S. Aydil, and
  L.~Gagliardi, ``Computational study of structural and electronic properties
  of lead-free csmi3 perovskites (m = ge, sn, pb, mg, ca, sr, and ba),'' {\em
  The Journal of Physical Chemistry C}, vol.~122, no.~14, pp.~7838--7848, 2018.

\bibitem{Eg_CGX4}
L.-C. Tang, C.-S. Chang, L.-C. Tang, and J.~Y. Huang, ``Electronic structure
  and optical properties of rhombohedral csgei3 crystal,'' {\em Journal of
  Physics: Condensed Matter}, vol.~12, p.~9129, oct 2000.

\bibitem{Eg_CGX5}
A.~J. Kale, B.~Pal, and A.~Dixit, ``Theoretical insights on pb-free
  rhombohedral csgei3 over cubic csmx3 (m-: Pb, sn, ge, and x-: Cl, br, i)
  perovskite-based single-junction solar cell with efficiency >30\%,'' {\em
  physica status solidi (a)}, vol.~221, no.~5, p.~2300464, 2024.

\bibitem{Eb2}
U.-G. Jong, C.-J. Yu, J.-S. Ri, N.-H. Kim, and G.-C. Ri, ``Influence of halide
  composition on the structural, electronic, and optical properties of mixed
  ${\mathrm{ch}}_{3}{\mathrm{nh}}_{3}\mathrm{Pb}{({\mathrm{I}}_{1\ensuremath{-}x}{\mathrm{Br}}_{x})}_{3}$
  perovskites calculated using the virtual crystal approximation method,'' {\em
  Phys. Rev. B}, vol.~94, p.~125139, Sep 2016.

\bibitem{Eb3}
A.~Miyata, A.~Mitioglu, P.~Plochocka, O.~Portugall, J.~T.-W. Wang, S.~D.
  Stranks, H.~J. Snaith, and R.~J. Nicholas, ``Direct measurement of the
  exciton binding energy and effective masses for charge carriers in
  organic–inorganic tri-halide perovskites,'' {\em Nature Physics}, vol.~11,
  2015.

\bibitem{WhenSOChelps}
T.~Das, G.~Di~Liberto, and G.~Pacchioni, ``Density functional theory estimate
  of halide perovskite band gap: When spin orbit coupling helps,'' {\em The
  Journal of Physical Chemistry C}, vol.~126, no.~4, pp.~2184--2198, 2022.

\bibitem{PhysRevB.93.195211}
L.-y. Huang and W.~R.~L. Lambrecht, ``Electronic band structure trends of
  perovskite halides: Beyond pb and sn to ge and si,'' {\em Phys. Rev. B},
  vol.~93, p.~195211, May 2016.

\bibitem{Zunger4}
C.~Mera~Acosta, E.~Ogoshi, A.~Fazzio, G.~M. Dalpian, and A.~Zunger, ``The
  rashba scale: Emergence of band anti-crossing as a design principle for
  materials with large rashba coefficient,'' {\em Matter}, vol.~3,
  pp.~2590--2393, Jul 2020.

\bibitem{VASP1}
G.~Kresse and J.~Hafner, ``Ab initio molecular dynamics for liquid metals,''
  {\em Phys. Rev. B}, vol.~47, pp.~558--561, Jan 1993.

\bibitem{VASP2}
G.~Kresse and J.~Furthmuller, ``Efficiency of ab-initio total energy
  calculations for metals and semiconductors using a plane-wave basis set,''
  {\em Comput. Mater. Sci.}, vol.~6, no.~1, pp.~15 -- 50, 1996.

\bibitem{VASP3}
G.~Kresse and J.~Furthm\"uller, ``Efficient iterative schemes for ab initio
  total-energy calculations using a plane-wave basis set,'' {\em Phys. Rev. B},
  vol.~54, pp.~11169--11186, Oct 1996.

\bibitem{VASP4}
G.~Kresse and D.~Joubert, ``From ultrasoft pseudopotentials to the projector
  augmented-wave method,'' {\em Phys. Rev. B}, vol.~59, pp.~1758--1775, Jan
  1999.

\bibitem{PBE}
J.~P. Perdew, K.~Burke, and M.~Ernzerhof, ``Generalized gradient approximation
  made simple,'' {\em Phys. Rev. Lett.}, vol.~77, pp.~3865--3868, 1996.

\bibitem{blochl94}
P.~E. Bl\"ochl, ``Projector augmented-wave method,'' {\em Phys. Rev. B},
  vol.~50, pp.~17953--17979, Dec 1994.

\bibitem{VASPKIT}
V.~Wang, N.~Xu, J.-C. Liu, G.~Tang, and W.-T. Geng, ``Vaspkit: A user-friendly
  interface facilitating high-throughput computing and analysis using vasp
  code,'' {\em Computer Physics Communications}, vol.~267, p.~108033, 2021.

\bibitem{KPOINTS}
H.~J. Monkhorst and J.~D. Pack, ``Special points for brillouin-zone
  integrations,'' {\em Phys. Rev. B}, vol.~13, pp.~5188--5192, Jun 1976.

\bibitem{pyprocar}
U.~Herath, P.~Tavadze, X.~He, E.~Bousquet, S.~Singh, F.~Muñoz, and A.~H.
  Romero, ``Pyprocar: A python library for electronic structure
  pre/post-processing,'' {\em Computer Physics Communications}, vol.~251,
  p.~107080, 2020.

\bibitem{Berry_phase}
R.~D. King-Smith and D.~Vanderbilt, ``Theory of polarization of crystalline
  solids,'' {\em Phys. Rev. B}, vol.~47, pp.~1651--1654, Jan 1993.

\bibitem{Eg_gap}
J.~P. Perdew and M.~Levy, ``Physical content of the exact kohn-sham orbital
  energies: Band gaps and derivative discontinuities,'' {\em Phys. Rev. Lett.},
  vol.~51, pp.~1884--1887, Nov 1983.

\bibitem{Zunger1}
C.~Mera~Acosta, L.~Yuan, G.~M. Dalpian, and A.~Zunger, ``Different shapes of
  spin textures as a journey through the brillouin zone,'' {\em Phys. Rev. B},
  vol.~104, p.~104408, Sep 2021.

\bibitem{Picozzi}
S.~Picozzi, ``Ferroelectric rashba semiconductors as a novel class of
  multifunctional materials,'' {\em Frontiers in Physics}, vol.~2, 2014.

\bibitem{Jagoda1}
K.~Tenzin, A.~Roy, H.~Jafari, B.~Banas, F.~T. Cerasoli, M.~Date, A.~Jayaraj,
  M.~Buongiorno~Nardelli, and J.~S\l{}awi\ifmmode~\acute{n}\else \'{n}\fi{}ska,
  ``Analogs of rashba-edelstein effect from density functional theory,'' {\em
  Phys. Rev. B}, vol.~107, p.~165140, Apr 2023.

\bibitem{REE2}
T.~S. Ghiasi, A.~A. Kaverzin, P.~J. Blah, and B.~J. van Wees, ``Charge-to-spin
  conversion by the rashba–edelstein effect in two-dimensional van der waals
  heterostructures up to room temperature,'' {\em Nano Letters}, vol.~19,
  no.~9, pp.~5959--5966, 2019.
\newblock PMID: 31408607.

\bibitem{CNG3}
R.~Kashikar, A.~Popoola, S.~Lisenkov, A.~Stroppa, and I.~Ponomareva,
  ``Persistent and quasipersistent spin textures in halide perovskites induced
  by uniaxial stress,'' {\em The Journal of Physical Chemistry Letters},
  vol.~14, no.~38, pp.~8541--8547, 2023.
\newblock PMID: 37724873.

\bibitem{CNG1}
R.~Kashikar, P.~S. Ghosh, S.~Lisenkov, A.~Stroppa, and I.~Ponomareva, ``Rashba
  effects in lead-free ferroelectric semiconductor
  $[{\mathrm{ch}}_{3}{\mathrm{ph}}_{3}{]\mathrm{SnBr}}_{3}$,'' {\em Phys. Rev.
  Mater.}, vol.~6, p.~104603, Oct 2022.

\end{thebibliography}
\end{document}